\documentclass[a4paper]{jpconf}
\usepackage{graphicx}
\begin{document}
\title{Field dependence of the thermopower of CeNiSn}

\author{$^1$U K\"ohler, $^1$P Sun, $^1$N Oeschler, $^2$T Takabatake, $^3$S Paschen and $^1$F Steglich}

\address{$^1$Max Planck Institute for Chemical Physics of Solids,
D-01187 Dresden, Germany}
\address{$^2$ADSM, Hiroshima University, Higashi-Hiroshima 739-8530, Japan}
\address{$^3$Institute of Solid State Physics, Vienna University of Technology,
1040 Vienna, Austria}

\ead{koehler@cpfs.mpg.de}

\begin{abstract}
Previously measured thermopower data of CeNiSn exhibit a significant
sample dependence and non-monotonous behavior in magnetic fields. In
this paper we demonstrate that the measured thermopower $S(T)$ may
contain a contribution from the huge Nernst coefficient of the
compound, even in moderate fields of 2~T. A correction for this
effect allows to determine the intrinsic field dependence of $S(T)$.
The observed thermopower behavior can be understood from Zeeman
splitting of a V-shaped pseudogap in magnetic fields.
\end{abstract}
\textbf{Introduction:} The orthorhombic system CeNiSn has been
classified as a Kondo semimetal, in which an anisotropic pseudogap
opens in the density of states (DOS) below approximately
10~K~\cite{CNS-95-4}.
Various experimental probes confirm the presence of a finite
quasiparticle DOS at the Fermi level, such as the metal-like
resistivities of samples with high purity~\cite{CNS-95-4} and the
linear-in-$T$ dependence of the thermal conductivity below
0.3~K~\cite{CNS-00-1}. Magnetic fields of the order of 10~T along
the easy magnetic $a$ axis suppress the gap formation significantly,
while fields along $b$ and $c$ are less effective
\cite{CNS-00-1,CNS-92-1}. The thermopower $S$ of CeNiSn is highly
anisotropic and exhibits a significant sample dependence
\cite{CNS-00-1,CNS-94-2,CNS-95-1,CNS-95-2}. Below 10~K, the absolute
values of $S$ are enhanced, which has been attributed to the gap
formation in this temperature range
\cite{CNS-94-2,CNS-95-1,CNS-95-2}. Consequently, application of a
magnetic field of 8~T along the easy $a$ axis has been found to
induce a significant lowering of the thermopower along $a$, $S_a$,
at low $T$~\cite{CNS-94-2,CNS-00-1}. However, the experimental data
for $S_b$ and $S_c$ with $B\parallel a$ are inconsistent. First
measurements at 4.2~K and 1.3~K showed a lowering of $S_c$ in
magnetic fields~\cite{CNS-95-1}, while investigations on samples of
higher purity revealed an increasing $S_c$ upon increasing
field~\cite{CNS-00-1}. Likewise, for $S_b$ at 1.3 K either a
monotonous decrease up to 10~T~\cite{CNS-95-1} or an increase in 4~T
with a subsequent decrease in 8~T~\cite{CNS-00-1} was found.

For correlated semimetals as CeNiSn a large Nernst coefficient has
been predicted~\cite{TB-07-2}. The corresponding transverse thermal
voltage can become comparable in magnitude to the longitudinal one
even in moderate magnetic fields. In such a case, small deviations
from the ideal contact geometry may give rise to a non-negligible
Nernst contribution to the measured thermopower, an effect which has
not been considered previously~\cite{CNS-95-1}. It is expected to be
most relevant for $S_b$, which exhibits significantly lower absolute
values at low temperatures than $S_a$ and $S_c$. In this work we
discuss the intrinsic behavior of $S_b(T)$ for $B\parallel a$ and
$c$ determined from
measurements in positive and negative fields up to 7~T. \\ \\
\textbf{Experiments:} The
investigated samples originate from the single crystal \#5, which
was grown by the Czochralski technique and subsequently purified by
the solid-state electrotransport (SSE) technique as described in
Ref.~\cite{CNS-95-4}. Thermopower and Nernst coefficient have been
measured with a steady-state method. The temperature gradient along
the samples was determined using a chromel-AuFe$_{0.07\%}$
thermocouple. In order to correct for spurious contributions due to
a non-ideal contact geometry all measurements were performed in $\pm
B$. Two configurations are presented: The heat current $q$ was
applied along $b$ with $B\parallel a$ (sample~1) and $B\parallel c$
(sample~2).\\ \\ \textbf{Results:}
\begin{figure}[!t]
\begin{center}
\includegraphics[width=0.96\textwidth]{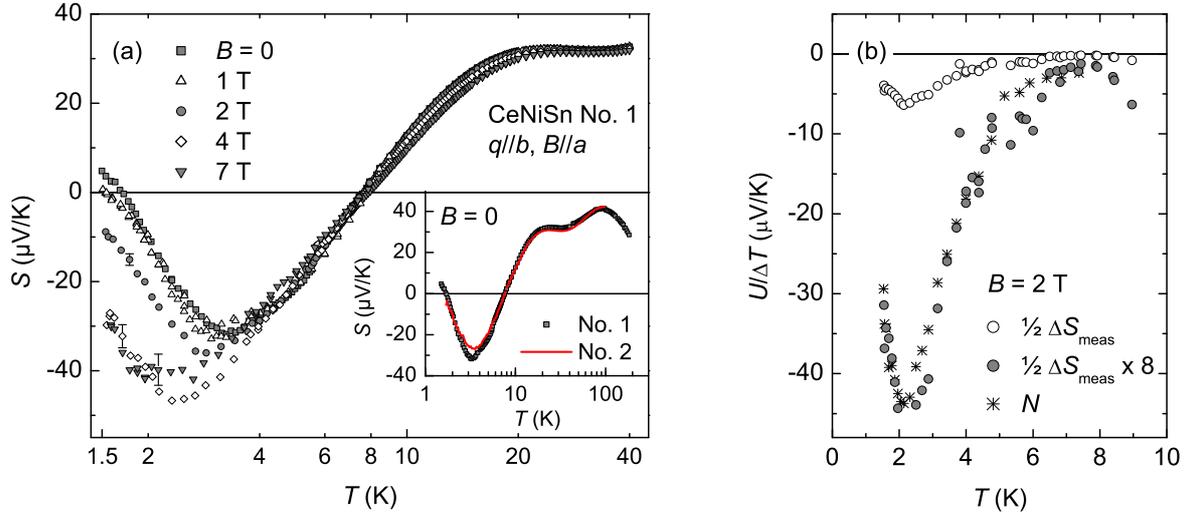}
\end{center}
\caption{(a) Thermopower of CeNiSn in different magnetic fields for
$q\parallel b$ and $B\parallel a$. Exemplary error bars indicate the
uncertainty of the data at low $T$ and high $B$. Above 4~K it is
typically less than 1~$\mu$V/K. The inset compares the zero-field
data obtained on two samples of CeNiSn. (b) The in-field
antisymmetric contribution of the measured thermopower $\frac{1}{2}
\Delta S_{\mathrm{meas}}$ in comparison to the Nernst signal $N$ in
2~T. $\frac{1}{2} \Delta S_{\mathrm{meas}}$ can be scaled to $N$ by
a factor of 8. \label{CNS-SvsT}}
\end{figure}
The zero-field thermopower of CeNiSn measured along $b$, $S_b(T)$,
is shown in the inset of Fig.~\ref{CNS-SvsT}. The data sets obtained
on two different samples agree relatively well in the whole
investigated temperature range. Above 10~K, the thermopower exhibits
a similar $T$ dependence as reported for a high-purity single
crystal without SSE treatment~\cite{CNS-95-2}: $S_b(T)$ is positive
with maxima at 20~K and 100~K. The two maxima are attributed to
Kondo scattering from the ground-state doublet and thermally
populated CEF levels. A contribution from paramagnon drag was also
suggested as a possible origin for the maximum at
20~K~\cite{CNS-00-1}. Toward lower $T$, $S_b (T)$ changes sign at
8~K and goes through a large negative minimum of $-30$ $\mu$V/K at
around 3.5~K. This value represents the largest negative $S$ ever
observed for a CeNiSn sample. Below 1.7~K the thermopower assumes
again positive values. A similar temperature dependence has been
observed in samples grown by a Czochralski technique without SSE
treatment~\cite{CNS-95-1}. These samples exhibit a negative $S_b$
between 2.5 and 7~K, however, with significantly smaller absolute
values of at most -6~$\mu$V/K. By contrast, investigations on single
crystals of similar quality as those presented here yielded a
thermopower $S_b(T)$ with opposite sign and a maximum value of
8~$\mu$V/K at 3.5~K~\cite{CNS-00-1}.

\begin{figure}[!t]
\begin{center}
\includegraphics[width=0.96\textwidth]{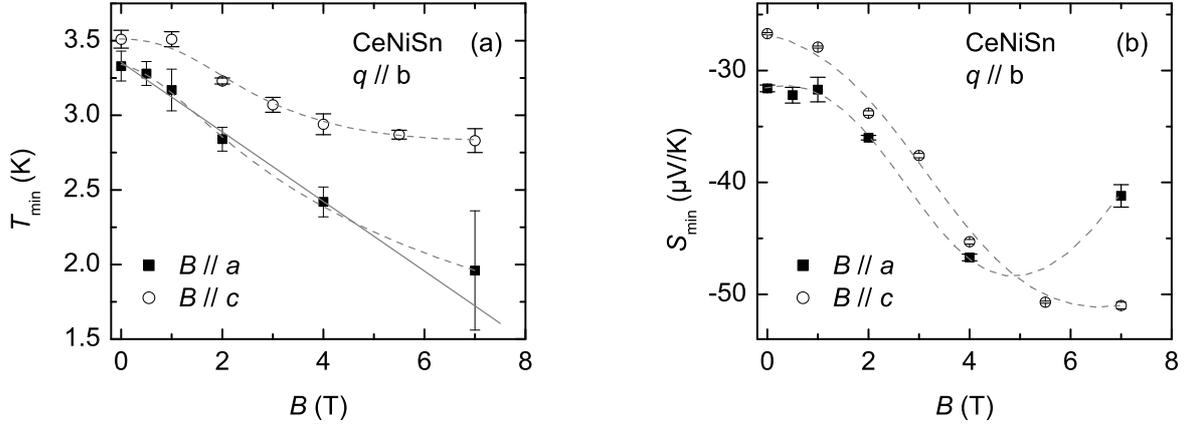}
\end{center}
\caption{Field dependence of the minimum temperature in $S_b(T)$,
$T_{\mathrm{min}}$ (a), and the thermopower value at
$T_{\mathrm{min}}$, $S_{\mathrm{min}}$ (b), for different
orientations of $B$. The dashed lines are meant guides to the eye.
The solid line in the left plot is a linear fit to the data for $B
\parallel a$. The error bars given for $S_{\mathrm{min}}$ represent
the scattering of the data around $T_{\mathrm{min}}$.
\label{CNS-Min_vsB}}
\end{figure}

The effect of a magnetic field along $a$ on the low-$T$ thermopower
$S_b$ is shown in the main plot of Fig.~\ref{CNS-SvsT}. With
increasing $B$ the minimum at $T_{\mathrm{min}}=3.5$~K shifts to
lower $T$ and the absolute values at the minimum
$|S_{\mathrm{min}}|$ are enhanced for $B \leq 4$~T. In 7~T a weak
lowering of $|S_{\mathrm{min}}|$ is observed, which, however, is of
the order of the uncertainty in the data. Application of a magnetic
field along $c$ gives rise to a similar evolution of $S(T)$ (not
shown). Compared to the configuration $B\parallel a$, the shift of
the minimum is less pronounced and no lowering of
$|S_{\mathrm{min}}|$ is observed around 7~T,
(cf.~Fig.~\ref{CNS-Min_vsB}). Fig.~\ref{CNS-Min_vsB}a shows the
field dependence of $T_{\mathrm{min}}$ for $B\parallel a$ and
$B\parallel c$. It clearly reveals that the shift of the minimum is
stronger for $B\parallel a$. A linear extrapolation of
$T_{\mathrm{min}} (B\parallel a)$ to zero temperature yields a
critical field of 14~T. This value is comparable to the energy-gap
quenching field along $a$ of 18~T determined from resistivity
measurements \cite{CNS-98-2}. Therefore, it is supposed that the
shift of the minimum is related to the closing of the gap in field.
Fig.~\ref{CNS-Min_vsB}b shows the evolution of the thermopower value
at the minimum, $S_\mathrm{min}$ for $B\parallel a$ and $B\parallel
c$. While $|S_{\mathrm{min}}|$ first increases with increasing
field, a saturation and subsequent reduction is observed at higher
$B$. Again, the effect of a field along $a$ is more pronounced than
that of $B
\parallel c$. It is suspected that $|S_{\mathrm{min}}|$ for both
orientations is further reduced in
higher $B$ as the minimum shifts to %
lower~$T$.\\ \\ \textbf{Discussion:} The strong sample dependence of
$S(T)$ of CeNiSn reported in literature has been related to the
differing purity of the investigated crystals~\cite{CNS-00-1}.
However, two other effects have to be taken into account. Firstly,
in view of the sensitive direction dependence of $S(T)$ it cannot be
excluded that tiny misorientations are responsible for at least part
of the reported variations. The large negative $S_b$ observed around
3~K in the present investigation could be easily diminished by a
small contribution from the huge positive $S_a$ and $S_c$ of up to
70~$\mu$V/K expected in the same temperature
range~\cite{CNS-00-1,CNS-94-2,CNS-95-1,CNS-95-2}. 
In this context, the small discrepancy of about 20~\% between
$S_{\mathrm{min}}(B=0)$ of sample 1 and 2 might be attributed to
this effect. Secondly, the large Nernst signal expected for CeNiSn
can contribute to the thermopower voltage for a non-ideal contact
geometry. In the present investigation, this effect was already
appreciable in 2~T as demonstrated in Fig.~\ref{CNS-SvsT}b. Below
8~K the measured thermopower curves $S_{\mathrm{meas}}(T)$ for
positive and negative fields differ significantly. The in-field
antisymmetric contribution $\frac{1}{2} \Delta S_{\mathrm{meas}} =
\frac{1}{2}[S_{\mathrm{meas}}(+2\mathrm{T})-S_{\mathrm{meas}}(-2\mathrm{T})]$
exhibits a temperature dependence similar to that of the Nernst
signal $N = E_y/\nabla_xT$ in 2~T and can be scaled to it by a
factor of 8. This corresponds to a misorientation of the contacts by
only $7^\circ$. For the presented data this effect has been
corrected for by averaging between $S_{\mathrm{meas}}(+B)$ and
$S_{\mathrm{meas}}(-B)$. However, it has generally not been
accounted for in previous investigations~\cite{CNS-95-1}. In
particular, the drastic and non-monotonous change in $S_b$ around
2~K reported for crystals of similar quality as those investigated
here~\cite{CNS-00-1} might be to some extent influenced by the huge
Nernst signal in this $T$ range. Therefore, the current study
represents the first investigation with the intrinsic temperature
dependence of $S_b$ in fields up to 7~T.

\begin{figure}[!tb]
\begin{center}
\includegraphics[bb=54 213 745 409,width=0.96\textwidth]{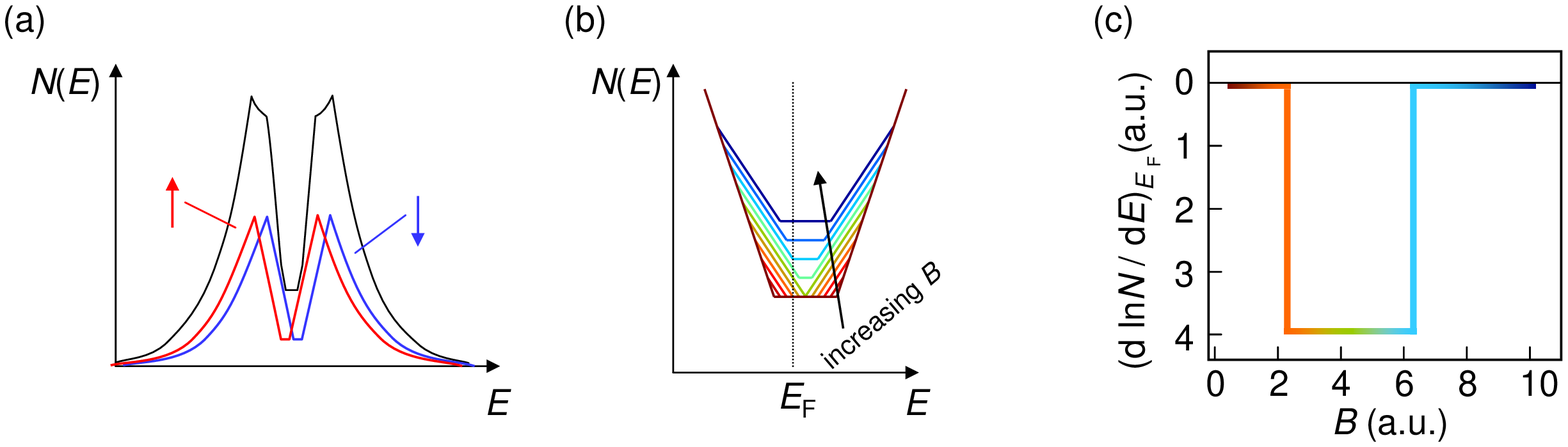}
\end{center}
\caption{(a) Schematic DOS of CeNiSn in a magnetic field according
to Ref.~\cite{CNS-97-2}. (b) Enhanced view of the region around
$E_{\mathrm{F}}$ for different $B$. (c) Field dependence of the
logarithmic derivative of the DOS at $E_{\mathrm{F}}$.
\label{CNS-Fig4x1}}
\end{figure}

Application of magnetic fields $B\parallel a, c$ is found to induce
a systematic shift of the thermopower minimum at 3.5~K to lower
temperatures (cf.~Fig.~\ref{CNS-Min_vsB}a). The estimated critical
field along $a$ of 14~T as well as the stronger effect for
$B\parallel a$ compared to $B \parallel c$ confirms that the minimum
is related to the gap formation in CeNiSn. In this context, the
enhancement of  $|S_\mathrm{min}|$ in parallel with the closing of
the gap in field (Fig.~\ref{CNS-Min_vsB}b) is surprising.
Apparently, application of a magnetic field does not only suppress
the gap but also influences the residual DOS inside the gap. A
similar effect has been found from investigations of the
specific-heat $c_p$ of CeNiSn \cite{CNS-97-2}, which revealed an
enhancement of $c_p$ at low $T$ in applied magnetic fields
$B\parallel a$. This observation had been interpreted within a
simple model assuming Zeeman splitting of a modified V-shaped
DOS~\cite{CNS-97-2}. An illustration, how the same mechanism can
give rise to an enhancement of $|S|$ at low $B$ is depicted in
Fig.~\ref{CNS-Fig4x1}: Application of a magnetic field induces a
shift of the sub-bands for the spin-up and spin-down states to
different energetic directions, as shown for an arbitrary field in
Fig.~\ref{CNS-Fig4x1}a. Thus, the resulting total DOS around the gap
structure depends sensitively on the field magnitude
(Fig.~\ref{CNS-Fig4x1}b). If the Fermi level is situated slightly
off the symmetry line of the gap structure, the slope of the DOS at
$E_\mathrm{F}$ becomes field dependent (Fig.~\ref{CNS-Fig4x1}c).
Within this simplified picture, the absolute values of the
thermopower $S \propto T(\partial \ln N(\epsilon)/\partial
\epsilon)_{E_{\mathrm{F}}}$~\cite{CNS-00-1} increase in small
magnetic fields and decreases for higher $B$. It is admitted, that
the sketched behavior is much too simple to explain in detail the
behavior of $S_b(T,B)$ observed for CeNiSn. 
Nevertheless, the presented picture demonstrates that the increase
in $|S|$ is not fundamentally contradictory to the suppression of
the gap in magnetic fields. Zeeman splitting appears a possible and
simple mechanism to understand the observed behavior. It seems
likely that a realistic band structure and the allowance for thermal
broadening enables a more sophisticated description of the data. In
conclusion, the minimum in $S_b(T)$ of CeNiSn at 3~K is attributed
to the gap formation and the observed field dependence is related to
the suppression of the gap in magnetic fields.

\end{document}